\documentclass[%
 aps,
 prl,
 superscriptaddress,
 amsmath,amssymb,
 aps,nofootinbib,   
]{revtex4}

\usepackage[a4paper,left=1.5cm,right=1.5cm,top=3cm,bottom=3cm]{geometry}

\usepackage{bm}
\PassOptionsToPackage{linktocpage}{hyperref}
\usepackage[hyperindex,breaklinks]{hyperref}
\usepackage{enumitem}
\usepackage{slashed}

\usepackage{float}
\usepackage{yfonts}

\usepackage{amsfonts}
\usepackage[dvipsnames]{xcolor}
\usepackage{graphicx}
\usepackage{longtable}
\usepackage{verbatim}
\usepackage{color}
\usepackage{mdframed}
\usepackage{soul}
\usepackage{amsfonts,amssymb,mathrsfs,amsmath,esint}
\usepackage{slashed, cancel}
\usepackage{framed}
\usepackage{mdframed}
\usepackage{simplewick} 
\allowdisplaybreaks

\usepackage{latexsym}
\usepackage{graphicx}
\graphicspath{{./Figures/}}
\usepackage[dvipsnames]{xcolor}
\usepackage{booktabs}
\usepackage{datetime}
\newdateformat{mydate}{\THEDAY{ }\monthname[\THEMONTH]{ }\THEYEAR}

\usepackage{tikz}
\usepackage{color}
\usepackage{framed}
\usepackage{hyperref}
\hypersetup{colorlinks, citecolor=bluscuro, linkcolor=black, urlcolor=bluscuro}
\definecolor{rossos}{cmyk}{0,1,1,0.55}
\definecolor{bluscuro}{rgb}{0.15, 0.2, .85}
\definecolor{bluchiaro}{cmyk}{1,.3,0.,0.1}

\newcommand{\ud}{{\rm d}}

\newcommand{\db}{\textswab{d}}

\newcommand{\be}{\begin{equation}}
\newcommand{\ee}{\end{equation}}
\newcommand{\ba}{\begin{eqnarray}}
\newcommand{\ea}{\end{eqnarray}}

\newcommand{\cH}{\mathcal{H}}
\newcommand{\bx}{{\bf x}}

\newcommand{\D}{\mathcal{D}}

\newcommand{\R}{\mathcal{R}}
\newcommand{\uM}{u_{\rm M}}
\newcommand{\thetaM}{\theta_{\rm M}}
\newcommand{\zetaM}{\zeta_{\rm M}}
\newcommand{\DM}{D_{\rm M}}
\newcommand{\cs}{c_{\rm s}}

\begin{document} 

\title{Breaking the Single Clock Symmetry: measuring single-field inflation non-Gaussian features.}

\author{Daniele Bertacca} 
\affiliation{Dipartimento di Fisica e Astronomia Galileo Galilei, Universit\`a di Padova, 35131 Padova, Italy}
\affiliation{INFN, Sezione di Padova, via F  Marzolo 8, I-35131 Padova, Italy.}
\affiliation{INAF - Osservatorio Astronomico di Padova, vicolo dell Osservatorio 5, I-35122 Padova, Italy.}
\email{daniele.bertacca@unipd.it}

\author{Raul Jimenez} 
\affiliation{ICC, University of Barcelona, Marti i Franques 1, 08028 Barcelona, Spain.}
\affiliation{ICREA, Pg. Lluis Companys 23, Barcelona, E-08010, Spain.}

\author{Sabino Matarrese}
\affiliation{Dipartimento di Fisica e Astronomia Galileo Galilei, Universit\`a di Padova, 35131 Padova, Italy}
\affiliation{INFN, Sezione di Padova, via F  Marzolo 8, I-35131 Padova, Italy.}
\affiliation{INAF - Osservatorio Astronomico di Padova, vicolo dell Osservatorio 5, I-35122 Padova, Italy.}
\affiliation{Gran Sasso Science Institute, viale F. Crispi 7, I-67100 L'Aquila, Italy.}

\author{Licia Verde}
\affiliation{ICC, University of Barcelona, Marti i Franques 1, 08028 Barcelona, Spain.}
\affiliation{ICREA, Pg. Lluis Companys 23, Barcelona, E-08010, Spain.}

\date{\today}

\begin{abstract}
The Universe is not just cold dark matter and dark energy, it also contains baryons, radiation and neutrinos. The presence of these components, beyond the pressure-less cold dark matter and the quasi-uniform dark energy ones,  imply that the single clock assumption from inflation is no longer preserved. Here we quantify this effect and show that the  single-clock symmetry is  ensured only on scales  where  baryonic effects, neutrinos effects, or sound speed are zero.  These scales depend on the cosmic epoch and the Universe composition.  Hence for all use and purposes of interpreting state-of-the-art and possibly forthcoming surveys, in the accessible scales, single clock symmetry cannot be said to be  satisfied. 
Breaking the single-clock symmetry has  key consequences  for the study of non-Gaussian features generated by pure single-field inflation which  arise from  non-linearities in the metric yielding  non-Gaussianities of the local type: the $n_{s}-1$ and the relativistic $-5/3$ term. \end{abstract}

\maketitle

\section{introduction}

The most economical model to describe the early Universe is a quasi-de Sitter state, i.e., inflation driven by a single field that is undergoing slow-rolling. Strong empirical support for inflation comes from two observational facts: the discovery of super-horizon fluctuations~\cite{WMAP03} and the near-scale-invariant spectrum of scalar fluctuations with a red tilt~\cite{WMAP03+,Planck:2018vyg}. In the simplest inflationary scenario, one single field dominates the quasi-exponential expansion of the space-time. This can happen because, at the energy scale of inflation, no other fields are light enough as to be able to provide enough e-folds on the expansion and are effectively frozen. This lack of other fields that are relevant during the inflation period indicates that one could describe inflation as a single clock with translational invariance of the field\footnote{For a description of the role of time during inflation see Ref.~\cite{Gomez:2021yhd}}. Under this symmetry, it is natural that one expects no coupling between the different fields and thus no deviation from Gaussian fluctuations as produced by the modes leaving the horizon without interaction. Self-coupling of the inflaton will produce a negligible non-Gaussian signal, of order ${\cal O}(\epsilon^2)$ where $\epsilon $ denotes the slow roll parameter. However, non-linearities of the single field paint a different story.

Non-linearities in the metric induced by the non-linearities of the single-field while slow rolling, give rise to two very well known effects~\cite{Gangui:1993tt,Gangui:1999vg,Wang:1999vf,Acquaviva:2002ud,Maldacena:2002vr}. The first is a non-Gaussianity contribution  with a  local  shape component  with  amplitude parameter $f_{\rm NL}^{\rm Local} \propto n_s - 1$  and an equilateral shape component of amplitude parameter  $f^{\rm equil.}_{\rm NL} \propto \epsilon$.
The second effect is  due to General Relativity and translates  effectively into a non-Gaussian contribution of the local type with $f_{\rm NL, eff.}^{\rm local}=-5/3$. 

 The inclusion of gravity in the analysis is crucial  as this non-Gaussian effect is generated by the non-linearities in the metric.  In Ref.~ \cite{Bellomo:2018lew}  a method was recently proposed to look at the effect of graviton interactions with scalar fluctuations of the inflaton field. This 
 opens the possibility to  study  observational signatures of single field inflation via the generation of gravitons, which is again a pure gravity effect that can only happen during a phase of accelerated expansion. 

Each of the above effects, primordial non-Gaussianity and relativistic effects, if measured,  would open  a new window into the physics of the early Universe. Despite some early claims that GR-effects driven non-Gaussian signatures  were simply gauge artifacts~\cite{Tanaka_2011,dePutter:2015vga,Pajer:2013ana,Cabass:2016cgp},  extensive work  e.g., ~\cite{Bertacca:2015mca, VerdeSasaki,Sasaki,Rollo}, has shown that indeed these non-Gaussianities are measurable and physical\footnote{One interesting and intuitive argument that shows that these higher order correlations cannot be gauged away is that if this was the case, one could also gauge away the 2-point correlation function (the power spectrum). This is very simply to see by using the Mukhanov-Sasaki variable in the in-in formalism with the proposed gauge transformations in Ref.~\cite{Tanaka_2011,dePutter:2015vga,Pajer:2013ana,Cabass:2016cgp}.}. In addition, it has been shown, from a quantum information theory point of view~\cite{QFNG} that single field non-Gaussianities cannot be gauge artifacts. Only in the unphysical case of exact $k=0$ can these non-Gaussianities be gauged away. 

In this work we take the calculation a step further and show that, in addition to the above arguments,  the presence in the Universe  of pressure(full) terms (e.g. baryons, radiation, neutrinos etc...)  do break the single clock symmetry and automatically imply that the above non-Gaussianities cannot be gauged away during the entire history  of the evolution of the patch i.e., on scales that are not significantly larger than the sound horizon at radiation drag.  This impossibility to synchronize clocks across cosmic epochs and separate patches, which size approaches the size of the current horizon,  has important consequences for the detectability of  primordial non-Gaussianity signatures.

\section{Methodology}

Let us start by considering a Universe with a Friedmann-Lema$\rm \hat i$tre-Robertson-Walker (FLRW) space-time at the background level, setting to zero the constant curvature $K$ (for simplicity), and assume General Relativity (GR) as the theory of gravity.
  The unperturbed energy-momentum tensor $T_{\mu \nu}$ of  matter contains ordinary (baryonic) matter, radiation,  neutrinos, Cold Dark Matter (CDM) and Dark Energy (DE) or, the cosmological constant $\Lambda$.  
At the background level, $T_{\mu \nu}$ describes a perfect fluid where the total energy density $\rho^{(0)}$, and the total pressure $p^{(0)}$ depend only on time~\footnote{The superscript $(0)$ denotes quantities at background.}.
The Einstein and continuity background equations are
\be\label{background}
\cH^2 = {8 \pi G \over 3} a^2  \rho^{(0)}\;,  \quad \quad \cH^2 - \cH' = 4 \pi G a^2 ( \rho^{(0)} +  p^{(0)}) \quad  {\rm and} \quad {\rho^{(0)}}'=-3\cH(p^{(0)}+\rho^{(0)})\;,
\ee
where  $\cH= a'/ a=a H$ is the comoving Hubble scale and prime denotes the derivative with respect to $\eta$. Here $\eta$ is the conformal time, i.e., $\eta = \int {\ud t / a(t)}$,
where $a$ is the scale factor. 
It is useful to define (see  Ref.~\cite{Mukhanov:1990me, Mukhanov:2005sc})  a  quantity obtained by the ratio between $ 0-0 $ and $ i-j $ components of Einstein's equations
\be \label{theta_M}
\theta_{\rm M} = {1 \over a} {1 \over \left(1 +  p^{(0)} /\rho^{(0)} \right)^{1/2}} ={ {\rho^{(0)}}^{1/2}\over a} \exp{\left[- {3 \over 2} \int^\eta (1+  \tilde c_s^2) \tilde {\cal H} \; \ud \tilde \eta \right]} \;,
\ee
that depends (at the functional level, i.e. $\theta_{\rm M} = \theta_{\rm M}[c_s]$) on the speed of sound $c_s^2={p^{(0)}}'/{\rho^{(0)}}'$.  Here we define two models: the spatially flat  $\Lambda$CDM model, which components are,  as it is customary, ordinary (baryonic) matter, radiation,  neutrinos, Cold Dark Matter (CDM) and the cosmological constant; we also define a pressureless counterpart of this model, the 0p$\Lambda$CDM model, which does not contain (or contains only a fully negligible contribution of) baryons, radiation and  neutrinos.   Hence the speed of sound just introduced is related to any possible effect that describes the physics beyond the pressure-less 0p$\Lambda$CDM model\footnote{Here we have defined $\tilde c_s^2 = c_s^2 (\tilde \eta)$ and $\tilde {\cal H} =  {\cal H} (\tilde \eta)$. Following Ref.~\cite{Mukhanov:2005sc}, the constant of integration in Eq.~(\ref{theta_M}) corresponds to an unphysical solution and, from now, it will be discarded.}. Using Eq.~(\ref{theta_M})  we note, immediately, that $ \cH^2 - \cH'  = (3/2) {\cal H}^2 / a^2 \theta_{\rm M}^2$.

We expand  the  background  around the pressureless case; using the bar to imply quantities in the  the corresponding 0p$\Lambda$CDM model, we have
 $\bar \cH^2 = 8 \pi G  \bar a^2  \bar \rho^{(0)}/3$, $\bar \cH^2 - \bar \cH' = 4 \pi G \bar a^2 (\bar \rho^{(0)} +  \bar p^{(0)})= 4 \pi G \bar a^2 \bar \rho^{(0)}_{\rm DM}$ and $ {{\bar \rho}^{(0)'}} = \bar \rho^{(0)'}_{\rm DM}  =-3\bar \cH(\bar p^{(0)}+\bar \rho^{(0)})=3\bar \cH \bar \rho^{(0)}_{\rm DM} $,
where $\bar \rho^{(0)} = \bar \rho^{(0)}_{\rm DM} + \bar \rho_\Lambda$, $\bar \rho^{(0)}_{\rm DM} \propto \bar a^{-3}$ and $\bar p^{(0)} = \bar p_\Lambda = - \bar \rho_\Lambda$. (Note that in general if today $a(\eta=\eta_0)=a_0=1$, then $\bar a_0 \neq 1$).  Of course, in a $\Lambda$CDM model with massless neutrinos,  one could adjust   the values of the model's parameters to ensure that $\Omega_{\rm CDM}$ in the 0p$\Lambda$CDM coincides with $\Omega_{\rm m}$ in the $\Lambda$CDM one, where baryons and cold dark matter are all included in the matter density parameter. In this case the scale factor evolution after recombination and for negligible amount of radiation, would effectively be the same  and $\bar{a}\simeq a$. Even in this hyper-simplified $\Lambda$CDM, $c_s$ is still non zero, but the effects can only be seen in the perturbations (see below) and not in the background.  However for non-massless neutrinos,  different neutrino masses transition from relativistic to non-relativistic at different times so  if  at a starting point, even if  well after matter radiation equality, we set  $a=\bar{a}$,  in general at later times  this will not hold in details. Moreover, if the dark energy were  not to be  a cosmological constant, the dark energy component could also contribute to $c_s$ in principle. For our analytical calculation, we have to consider the initial conditions, which we denote with subscript (in). We set initial conditions in the matter dominated era,  and we consider two cases, one  well before recombination and one  well after recombination. 

In addition, at  the background level (in FLRW space-time), we assume that  the parameter expansion $a$ at $\eta_{\rm in}$ coincides  in the 0p$\Lambda$CDM and $\Lambda$CDM model\footnote{In principle we could use an alternative condition,  i.e.,  the parameter expansion could coincide  both for 0p$\Lambda$CDM and $\Lambda$CDM model today at $z=0$. In the main text we have used the condition during matter epoch in order to match our considerations with approach used in separate universe technique. See also the discussion at the end of the next section.}. 

For a generic quantity, let us call  it $C^{(0)}$  which in this section we take to be a background quantity,  we define, at the linear level,  the deviations from 0p$\Lambda$CDM  due to the presence of a small pressure component as
\be
 \db C^{(0)} = C^{(0)} - \bar C^{(0)} \;, \quad  {\rm where} \quad  \left|\db C^{(0)} /\bar C^{(0)} \right| \ll 1\,.
\ee

Note that for the pressure component to be small, the size of the patch of the Universe under consideration needs to be much larger than the largest scale reached by the sound horizon (the Jean's length) from the initial conditions on.  Hence, for initial conditions set before recombination,  that corresponds to scales much larger than the sound horizon at radiation drag. If we instead set initial conditions after recombination, this scales needs to be larger than  massive neutrinos free streaming length, or  for massless neutrinos, larger than the scale where baryonic effects become important. Recall that the sound horizon at radiation drag is $\sim 150$ Mpc, the neutrinos free streaming length corresponds to $k\sim 0.01 h$ Mpc$^{-1}$, and baryonic effects are important on scales of few Mpc. 

 Immediately we note that  $ \db C^{(0)}$  depends on $c_s^2$ 
and it is possible to identify explicitly $ \db  p^{(0)} $ from the definition of the speed of sound. Indeed we obtain
\be \label{deltabarp}
 \db  p^{(0)} [c_s^2]=  \db  p^{(0)}_{\rm (in)}+ \int_{\eta_{\rm (in)}}^{\eta} \tilde c_s^2 {\tilde \rho}^{(0)'} ~\ud \tilde \eta. 
\ee

From  $a = \bar{a} + \db a$ 
\be
\cH^2=\bar \cH^2\left(1+2{\db \cH \over \bar \cH}\right)={8\pi G \over 3}\bar a^2\bar \rho^{(0)} \left(1 + {\db a \over \bar a}\right)\left(1+ {\db \rho^{(0)} \over \bar \rho^{(0)}}\right)
\ee
and
\be
\cH=(\bar \cH + \db \cH)= \left(1 - {\db a \over \bar a}\right)\left(\bar \cH+{\db a'\over \bar a}\right)\;,
\ee
we have the following relations for $\db \cH$
\ba \label{deltabarcH1}
\db \cH =  {\bar \cH \over 2} \left({\db \rho^{(0)} \over \bar \rho^{(0)}}+2{\db a \over \bar a}\right)\;, \quad \quad
\db \cH= {\db a' \over \bar a}-\bar \cH {\db a \over \bar a}
\ea
or, equivalently, combining the above relations 
\ba
\label{cH} 2\db \cH= {\db a' \over \bar a} +{\bar \cH \over 2}{\db \rho^{(0)} \over \bar \rho^{(0)}}\;,\quad \quad
\label{da/a}{\ud \over \ud \bar a}\left(\db a \over \bar a^2 \right)={1\over 2\bar a^2}{\db \rho^{(0)} \over \bar \rho^{(0)}}\;.
\ea
From the continuity equation
we find
\be\label{pert-background_continuity_eq}
{\db \rho^{(0)}}'+3\bar \cH\left(\db  p^{(0)}+ \db \rho^{(0)} \right)=-3 (\bar p^{(0)} + \bar \rho^{(0)}) \db \cH \,.
\ee
Here we note that as long as we are only concerned with times well after neutrino decoupling,  it is possible to go from the 0p$\Lambda$CDM model to the $\Lambda$CDM model in a continuous and differentiable way, i.e. the background quantities $C$ are continuous and differentiable.

This equation, using Eq. (\ref{cH}),  can be rewritten as
\be \label{diffeqdeltabarrho}
\frac{\ud^2 \db \rho^{(0)}}{\ud \bar a^2}+{3\over  2\bar a}(\bar w+5)\frac{\ud \db \rho^{(0)}}{\ud \bar a}+ {3\over  2\bar a^2} [4+3(\bar w +1)^2]\db \rho^{(0)}+{6\over  \bar a^2}   \db  p^{(0)}+ {3\over  \bar a} \frac{\ud \db  p^{(0)}}{\ud \bar a}=0\;,
\ee
where we defined $\bar w=\bar p^{(0)}/\bar \rho^{(0)}$. This is the complete and correct equation to  solve in the generic case. However it has no analytic solution and is therefore not too transparent. We proceed by considering a symplified case, which, however, still captures all the interesting aspects. For example  in the  $\Lambda$CDM case  we can directly get  $\db \rho^{(0)}=  (\rho^{(0)} -  \bar \rho^{(0)} )$ from the Friedmann equation
  
 \ba
\rho^{(0)}(z) &=& { 3 H_0^2 \over 8 \pi G } \Bigg\{
\left(\Omega_{{\rm CDM}0}+\Omega_{{\rm b}0}\right)(1+z)^3+\Omega_{\gamma0}(1+z)^4
\left[1+0.2271N_{\rm eff} f\left({m_\nu \over [(1+z)T_{\nu
 0}]}\right)\right]\nonumber\\
 &&+ \Omega_{\Lambda0}
\Bigg\}, \quad    {\rm and} \quad    \bar \rho^{(0)}(z) =  { 3 \bar H_0^2 \over 8 \pi G } \Big[ \Omega_{{\rm CDM}0}(1+z)^3 +  \Omega_{\Lambda0} \Big]\;,
\label{eqHubblegeneral}
\ea
where $ f$ is a suitable function in which $(\Omega_\nu/a^4)  0.2271N_{\rm eff} f(m_\nu a/ T_{\nu 0}) \to \Omega_\nu/a^3$ for $a \to \infty$ (e.g. see \cite{WMAP:2010qai}). In the massless neutrinos case $f=1$.
Here the values of  different  cosmological parameters  for the $\Lambda$CDM model are constrained in Ref.~\cite{Planck:2018vyg} and for the 0p$\Lambda$CDM the reader can think of absorbing the baryonic component into the cold component to keep $\Omega_m$ the same. 

In general, if we choose $\db a_{\rm (in)} =0$, 
using Eq. (\ref{deltabarcH1}), we obtain
\be
 \db \cH_{\rm (in)}  =  {\bar \cH_{\rm (in)}  \over 2} {\db \rho^{(0)}_{\rm (in)}  \over \bar \rho^{(0)}_{\rm (in)} }\;.
\ee
Using  Eq.~(\ref{da/a}), we find
\be
\left({\db a \over \bar a^2 }\right)
={1\over 2} \int^{a}{1\over \tilde a^2}{\db \tilde \rho^{(0)} \over \tilde \rho^{(0)}} \ud \tilde a \;,
\ee
and, taking into account Eq.~(\ref{pert-background_continuity_eq}), we have
\be
 \db  p^{(0)}_{\rm (in)} = - {(3+\bar w_{\rm (in)})\over 2} \db \rho^{(0)}_{\rm (in)} - {\db \rho^{(0)'}_{\rm (in)} \over 3 \bar \cH_{\rm (in)} }\;.
\ee

Finally, from Eq. (\ref{theta_M}), the correction for $\theta_{\rm M}$ becomes 
\be
 \db \theta_{\rm M} = \bar \theta_{\rm M} \left[{1\over 2} \left({\db \rho^{(0)} \over \bar \rho^{(0)}} - {\db p^{(0)} + \db \rho^{(0)} \over \bar p^{(0)}+ \bar \rho^{(0)}} \right)- {\db a \over \bar a } \right]
\ee
where
\be \label{bar-theta_M}
\bar \theta_{\rm M} = {1 \over \bar a} {1 \over \left(1 + \bar  p^{(0)} /\bar \rho^{(0)} \right)^{1/2}} =  {1 \over \bar{a}} 
\left({\bar \rho^{(0)} \over {\bar \rho}_{\rm CDM}^{(0)} }\right)^{1/2}\;.
\ee
Here ${\bar \rho}^{(0)}_{\rm CDM}$ denotes the density of cold dark matter and we have assumed that for dark energy $p=-\rho$.

 While at the background level  baryons,  radiation and neutrinos (massless, fully relativistic or  massive and non-relativistic) have a negligible effect, it is important to bear in mind  that at the perturbation level massive neutrinos  and baryons have important effects even at linear scales. Baryons are known to  have effects  on the mildly non-linear regime (the highly feared baryonic effects) which, however at linear scales  are at the few percent level. Massive neutrinos on the other hand,  have non zero velocity dispersion which acts as an effective sound speed; because of this, they have effects  on the power spectrum at scales $k>0.01 h$ Mpc$^{-1}$ comparable to the  free streaming length; the effect is scale- and redshift-dependent. Even in the linear regime, at $k\sim 0.1  h$ Mpc$^{-1}$  neutrinos produce a suppression on the power spectrum of $\sim 6f_{\nu}$ where $f_{\nu}$ denotes the neutrino mass fraction\footnote{For the minimum neutrino mass allowed by oscillation experiments, and for the concordance cosmology, $f_{\nu}$ is $\sim 0.01$.}. At the minimum neutrino mass imposed by oscillations experiments, the linear power spectrum  suppression is $\sim 6\%$ at $z=0$. It is well known that  massive neutrinos also introduce a  scale-dependent  bias~\cite{loverdenu} and affect peculiar velocities. Hence their effects likely cannot be neglected in the perturbations. In particular while in the absence of massive neutrinos the matter density perturbations in the linear regime evolve $\propto a$,  in the presence of massive neutrinos evolve as $a^{1-3/5f_{\nu}}$. 

\subsection{Corrections on cosmological perturbations}
\label{sec:perturbations}
Consider the correction in the scalar metric perturbations in the longitudinal gauge  \cite{Bardeen:1980kt, Kodama:1985bj, Bertschinger:1993xt, Malik:2008im} (or also called Conformal Newtonian Gauge \cite{Mukhanov:1990me, Ma:1995ey, Mukhanov:2005sc}). In this gauge\footnote{Here $\Phi=\Phi_{A}Q^{(0)}$ and $\Psi=- \Phi_{ H}Q^{(0)}$, where $\Phi_{A}$ and $\Phi_{ H}$ are the well known Bardeen potentials \cite{Bardeen:1980kt}.}
 \begin{eqnarray} 
 \label{Long-metric}
 \ud s^2 &=& a(\eta)^2\left[-\left(1 + 2\Phi \right)\ud\eta^2+ \delta_{ij} \left(1 -2\Psi \right) \ud x^i\ud x^j\right] .
\end{eqnarray}

Let us introduce the following variables (already used in \cite{Mukhanov:1990me, Mukhanov:2005sc}):\footnote{The constants of integration arising in these formulae correspond to unphysical solution that can be removed.}
\ba\label{uM}
\uM&=&\exp\left[{3\over 2} \int^\eta \ud \tilde \eta \; \left(1+\tilde c_s^2\right) \tilde \cH \right] \Psi =  {a \thetaM \over \left(\rho^{(0)}\right)^{1/2}} \Psi\;, \\
\label{zetaM}
\zetaM&=& \Psi - {\cH^2 \over \cH' - \cH^2} \left( \Psi + {\Psi' \over \cH} \right)={2 \over 3} \left({3 \over 8 \pi G}\right)^{1/2} \thetaM^2 \left({\uM \over \thetaM}\right)'\;,
\ea
and the comoving curvature perturbation that can be written in terms of $\zetaM$ and $\DM$
\be\label{calR}
\R= \Psi - {\cH^2 \over \cH' - \cH^2} \left( \Phi + {\Psi' \over \cH} \right) = \zetaM + {2 \over 3} a^2 \thetaM^2 \DM\;,
\ee
where $\DM =\Phi - \Psi $. (Of course, in 0p$\Lambda$CDM model, we have $\DM=0$ and $\R$ coincides with $\zetaM$).
Using the linear perturbation of the spatial part of the stress energy tensor 
\[ T^{i (1)}_j = p^{(1)} \delta^i_j + \D^i_j \Pi \;,\]
where $ \D_{ij}=\partial_i \partial_j - \delta_{ij} \nabla^2/3 $, $p^{(1)}$ is the perturbation of isotropic pressure\footnote{Here $p^{(1)}$  is not the same  $\delta p$ defined in \cite{Mukhanov:1990me}.} and $\Pi$ is the trace-free scalar part of total anisotropic stress tensor \cite{Malik:2008im}, thorough
 the Einstein equation we can write 
\be \label{Pi}
\DM = -8\pi G a^2 \Pi \;.
\ee
Finally, using (\ref{zetaM}), Eq.~(\ref{calR}) becomes
\be\label{calR2}
\R=\zetaM - {16 \pi G \over 3} a^2 {\rho^{(0)} \over \left(p^{(0)}+\rho^{(0)} \right) }\Pi\;.
\ee
Deviations from the isotropic pressure can be split in the adiabatic perturbations (which is proportional to $\cs^2$) and the intrinsic non-adiabatic pressure perturbation\footnote{Note that $p^{(1)}$ is the pressure perturbation in the longitudinal gauge.} \cite{Bardeen:1980kt, Kodama:1985bj}  where $\Gamma$ is the non-adiabatic component of the equation of state and the $\delta$ is the adiabatic component.
\be
p^{(1)} = p^{(0)} \left(\Gamma + {c_{\rm s}^2 \over w} \delta \right)\;,
\ee
and we can easily generalise  the prescription made in Ref.~\cite{Mukhanov:1990me, Mukhanov:2005sc}. Indeed,  with some algebra, we can  obtain
\be
\left[ \thetaM^2 \left({\uM \over \thetaM}\right)' + \left(8\pi G \over 3\right)^{1/2} a^2 \theta^2 \DM \right]'=\cs^2 \thetaM^2 \nabla^2 \left({\uM \over \thetaM}\right) - { a^2 \thetaM^2 \over 3 {\rho^{(0)}}^{1/2}} \nabla^2 \DM + 4 \pi G {a^3 \thetaM^2 p^{(0)} \over  {\rho^{(0)}}^{1/2}} \Gamma\,.
\ee
Without $D_M$ and $\Gamma$ this is identical to the results of \cite{Mukhanov:2005sc}, hence this is the generalization of that calculation in presence of pressure terms.

Using Eqs.~(\ref{uM}), (\ref{zetaM}), (\ref{calR2}), (\ref{Pi}) and the gauge-invariant definition of the Newtonian Poisson equation 
\be \label{NewtonianPoisson}
\nabla^2 \Psi = 4 \pi G a^2 \rho^{(0)} \Delta_{\rm com} = 4 \pi G a^2  \rho^{(1)}_{\rm com}\;,
\ee
where $\Delta_{\rm com}$ is the density contrast and $ \rho^{(1)}_{\rm com}$ is density  perturbation in the comoving orthogonal gauge, we obtain the usual relation 
\be \label{calRprim}
{ \left(p^{(0)}+\rho^{(0)} \right) \over \cH}\R'= c_{\rm s}^2 \, \rho^{(1)}_{\rm com} + {2 \over 3} \nabla^2 \Pi + p^{(0)} \Gamma\;.
\ee
The time evolution of $\R$ in a spatially flat background is dictated by three components: the sound speed modulated by the comoving density perturbation,  the spatial variation (second derivative) of the shear and non-adiabatic pressure perturbations: $\R'=0$ on scales and epochs where $c_{\rm s}$, $\Gamma$ and  $\nabla^2 \Pi$ are negligible. Before recombination this happens on scales $\gg $ 150 Mpc, after recombination depending on the neutrino mass the scales are  well above  tens  or few megaparsecs (respectively). 

Some other important comments are in order here:
\begin{itemize}
\item
Due to the breaking of the Single Clock Symmetry, for each point in space we have a non zero acceleration due to  $c_s$, $\Pi$ and/or $\Gamma$ and we cannot build an exact observers proper reference frame described by the local (normal) coordinate within the patch  along all the observer world line. 
\item
If we wanted to build a local volume expansion it is necessary to describe the physics inside this patch with the coordinates of an observer that are comoving with matter. For this purpose, the comoving gauge is the most appropriate.

In this gauge the curvature potential is $\R$ and the lapse perturbation are described by the $00$ metric perturbation $\xi=g_{00}^{(1)}|_{\rm com}$ which can be written in terms of pressure perturbation and anisotropic stress
\be
\left(p^{(0)}+\rho^{(0)} \right) \xi =   c_{\rm s}^2 \, \rho^{(1)}_{\rm com} + {2 \over 3} \nabla^2 \Pi + p^{(0)} \Gamma\;.
\ee
Then, using Eq.~(\ref{calRprim}), we have $\xi= \R' / \cH$.
We note immediately that  we can follow the separate Universe prescription (e.g., see also \cite{Hu:2016wfa}) and using the above equations  we are able to compute  the curvature evolution equation within the patch. We find \cite{Hu:2016wfa} 
\be
{ K_{\rm patch}' \over \cH}= -{2\over 3} \nabla^2 \xi =  -{2\over 3 \left(p^{(0)}+\rho^{(0)} \right) } \nabla^2 \left[ c_{\rm s}^2 \, \rho^{(1)}_{\rm com} + {2 \over 3} \nabla^2 \Pi + p^{(0)} \Gamma\right]\;.
\ee

Note that $K_{\rm patch}$ is not conserved if there is flow of material in or out of the patch. For example, for a patch of $5$ Mpc corresponding roughly  to the Lagrangian radius of a massive elliptical, non-linear evolution would yield a lot of material to enter this radius through the  surrounding filaments, even for a 0p$\Lambda$CDM Universe.
Therefore the curvature $K_{\rm patch}$ is conserved if the size of the patch is large enough and/or for $p^{(1)}_{\rm com}$ and $\Pi$ equal to zero, i.e. in the 0p$\Lambda$CDM model. Consequently, if  $K_{\rm patch}$ is not conserved  the lapse perturbation is not zero. This is another way to see that the single clock symmetry  is broken.
\item As an example, it is useful quantify the contribution of  $\Pi$. For example the well-know effect due to neutrinos is
\be
D_{\rm M}= - {2 \over 5} R_\nu \Phi= - {2 R_\nu /5 \over 1 + 2 R_\nu /5 } \Psi\;,
\ee
where \cite{WMAP:2010qai}
\be
R_\nu = {0.2271N_{\rm eff} f\left({m_\nu a / T_{\nu 0}}\right) \over 1 + 0.2271N_{\rm eff} f\left({m_\nu a / T_{\nu 0}}\right)} \;.
\ee
Then using  the Newtonian Poisson equation Eq.~(\ref{NewtonianPoisson}) we find
\be
{2 \over 3} \nabla^2 \Pi = {1 \over 3}\left( {2 R_\nu /5 \over 1 + 2 R_\nu /5 }\right) \rho_{\rm com}^{(1)}
\ee
and we immediately note that the the single clock is explicitly broken due to the neutrinos 
\be
\left(p^{(0)}+\rho^{(0)} \right) \xi =   \left[ c_{\rm s}^2 
+  {1 \over 3}\left({2 R_\nu /5 \over 1 + 2 R_\nu /5 }\right) \right] \rho_{\rm com}^{(1)} + p^{(0)} \Gamma\;.
\ee 
\end{itemize}
These points have a fundamental consequence. In order to quantify the size of 
the patch which preserves the curvature $K_{\rm patch}$ 
at a given  point of the space we have to consider the entire evolution of the separate Universe  from $a_{\rm in}$, until today. (Note that our initial condition can be chosen so that $a_{\rm patch}|_{\rm in}=a_{\rm in}$, where $a_{\rm patch}$ is the local scale factor of the patch local Universe, see also ~\cite{Hu:2016wfa}).
The patch radius $R_{\rm patch}$ enclosing constant mass  has to be above the Jeans scale along all the observer’s world line. For example, assuming massive neutrino, we should take scales with $k < k_{\rm nr}$. Here, for  individual neutrino masses  $ m_\nu$ from $0.046$ to $0.46$ eV, the scale $k_{\rm nr}$  ranges from $2.1 \times 10^{-3}h$Mpc$^{-1}$ to $6.7\times10^{-3}h$ Mpc$^{-1}$ \cite{Lesgourgues:2012uu} (see also \cite{garay,TopicalConvenersKNAbazajianJECarlstromATLee:2013bxd}).  Of course at recombination the Jeans length for baryons is  the sound horizon which is comparable to the horizon. This point confirms what was previously discussed earlier, i.e.  these  patches are the only places where clocks can be synchronized, but they are the ones with the most cosmic variance.

In the next subsection we analytically calculate the corrections that break the validity of the separate Universe prescription (see also Ref.~\cite{Hu:2016wfa}).
As we will see, this implies that the single clock from inflation is no longer preserved.

\subsubsection{Correction to linear order perturbations}

Denoting again the concordance model quantities with an over-bar, 
we split  the linear order perturbations, e.g. let us call $C^{(1)}$, in the following way
\be
C^{(1)} = \bar C^{(1)} + \db C^{(1)} \;,
\ee 
where
\ba
\db C^{(1)} =  \int {\db \rho^{(0)}}  \left({\delta  C^{(1)} \over \delta {\db \rho^{(0)} } }\right) \ud \tilde \eta  &+& \int \cs^2  \left({\delta  C^{(1)} \over \delta \cs^2}\right) \ud \tilde \eta \nonumber\\
&+&  \int {\Gamma}   \left({\delta  C^{(1)} \over \delta \Gamma}\right) \ud \tilde \eta \, \ud^3 \tilde {\bf x} +  \int {\Pi}   \left({\delta  C^{(1)} \over \delta {\Pi}}\right) \ud \tilde \eta \, \ud^3 \tilde {\bf x} + {\rm ...}\,. \nonumber\\ 
\ea
Here, for simplicity,  we are assuming that the correction, i.e.,  the  contribution due to $\{\db \rho^{(0)} = \rho^{(0)} - \bar \rho^{(0)},  \cs^2, \Gamma, \Pi\}$,  is  only linear. In principle, it is easy generalise this point. Immediately we note that from Eq.~(\ref{calRprim}) we have $\bar \R' =0$,
\ba \label{dbR}
 \db \R = \db \R^{\rm (in)} +{3 \over 2}&& \left({3 \over 8 \pi G}\right)^{1/2} \int_{\rm in} \tilde \cs^2   \bar \theta_{\rm M}(\tilde \eta)  \nabla^2 \bar u_{\rm M}(\tilde \eta,  {\bx}) \, \ud \tilde \eta \nonumber\\
&&+ \int_{\rm in} {\cH (\tilde \eta)\over \left(\bar p^{(0)}(\tilde \eta)  + \bar \rho^{(0)}(\tilde \eta)\right) } \Bigg[ {2 \over 3 }\nabla^2 \Pi(\tilde \eta,  {\bx})  + \bar p^{(0)} \Gamma  (\tilde \eta,  {\bx}) \Bigg] \ud \tilde \eta 
\ea
and, using Eqs.~(\ref{calR}) and (\ref{calR2}), we obtain
\be
\bar \zeta_{\rm M} = \bar \R \quad \quad {\rm and} \quad \quad \db \zeta_{\rm M} =  \db \R  - {2 \over 3} {\bar \rho^{(0)} \over \left(\bar p^{(0)}+\bar \rho^{(0)}   \right) } \DM =  \db \R  + {16 \pi G \over 3} \bar a^2 {\bar \rho^{(0)} \over \left(\bar p^{(0)}+\bar \rho^{(0)} \right) }\Pi\;.
\label{eq:master}
\ee

The leading contribution to $\Pi$ depends again  on the scale, the  cosmic epoch and the Universe composition.  Before recombination is dominant inside the horizon.  After recombination, for massive neutrinos,  free streaming is not negligible even on linear scales;  for baryons, it is important on cluster scales and below. 

We note that Eq.~(\ref{eq:master}) is our main result, and at the end of this section  we compute the initial conditions. It is transparent that it is not possible to completely eliminate the $c_s$ dependence from the perturbed variable, except in the (academic) case of infinite wavelength $k=0$.
\\

Now, splitting  each term in Eq.~(\ref{zetaM}), we  find immediately the solution for $\db u_{\rm M}$:
\ba
 \bar u_{\rm M} &=&  {3 \over 2}\left({3 \over 8 \pi G}\right)^{1/2} \bar \theta_{\rm M}  \int_{\rm in} {\bar \zeta_{\rm M} (\tilde \eta,  {\bx}) \over \bar \theta^2_{\rm M} ( \tilde \eta )} \ud \tilde \eta +  {\bar \theta \over \bar \theta_{\rm M}^{\rm (in)}  }   {\bar u_{\rm M}}^{\rm (in)} \;,   \quad {\rm for }~\Lambda{\rm CDM ~ model,~ ~ and}\nonumber \\
 \db u_{\rm M}&=& {\db  \theta_{\rm M}\over \bar \theta_{\rm M}} \bar u_{\rm M} +  {3 \over 2}\left({3 \over 8 \pi G}\right)^{1/2} \bar \theta_{\rm M}  \int_{\rm in} \left( - 2 {{\db \theta_{\rm M}} (\tilde \eta) \over \bar \theta_{\rm M}^3 (\tilde \eta)  }   \bar \zeta_{\rm M} (\tilde \eta, \bx)+ {\db \zetaM (\tilde \eta, \bx)\over \bar \theta_{\rm M}^2 (\tilde \eta) }  \right) \ud \tilde \eta
 +\bar \theta_{\rm M} \Bigg(  { {\db u_{\rm M}}^{\rm (in)} \over {\bar \theta_{\rm M}}^{\rm (in)} }- { {\db \theta_{\rm M}}^{\rm (in)} \over \left({\bar \theta_{\rm M}}^{\rm (in)} \right)^2 }  {\bar u_{\rm M}}^{\rm (in)}  \Bigg) \;.\nonumber \\
 \ea
From Eq.~(\ref{uM}), we get
\be
\bar \Psi = { \left(\bar \rho^{(0)}\right)^{1/2}  \over \bar a \bar \thetaM}  \bar u_{\rm M} = \left(\bar p^{(0)} + \bar \rho^{(0)}  \right)^{1/2} \bar u_{\rm M} \quad  \quad {\rm and} \quad \quad \db \Psi = \left(\bar p^{(0)} + \bar \rho^{(0)}  \right)^{1/2} \db u_{\rm M} + {1 \over 2 }{\db p^{(0)} + \db \rho^{(0)} \over  \left(\bar p^{(0)} + \bar \rho^{(0)}  \right)^{1/2} } \bar u_{\rm M} \;,
\ee
and 
from Eq.~(\ref{Pi}), we obtain
\be \label{dbPhi}
\db \Phi = \db \Psi + \DM =  \db \Psi  -8\pi G \bar a^2 \Pi\;.
\ee
Finally from the Newtonian Poisson Eq.~(\ref{NewtonianPoisson}) we deduce
\be
\bar \Delta_{\rm com} = {1 \over 4 \pi G \bar a^2 \bar \rho^{(0)}  }\nabla^2  \bar \Psi  \quad  \quad {\rm and} \quad \quad  
\db \Delta_{\rm com}   =  {1 \over 4 \pi G \bar a^2 \bar \rho^{(0)}  }\nabla^2  \db \Psi - \left( 2 {\db a \over \bar a}  + {\db \rho \over \bar \rho^{(0)}} \right)\bar \Delta_{\rm com}\;.
\ee

We suggest  to set the initial conditions by the following approach. From $\db a_{\rm in}=0$, we can choose $\db \Psi^{\rm (in)}  \equiv \Psi^{\rm (in)}  - \bar \Psi^{\rm (in)}  $ where $\Psi^{\rm (in)} =  4 \pi G \bar a_{\rm (in)}^2  \rho^{(0)}_{\rm (in)} \nabla^{- 2} \Delta_{\rm com}^{\rm (in)} $ and $\bar \Psi^{\rm (in)}  =  4 \pi G \bar a_{\rm (in)}^2 \bar \rho^{(0)}_{\rm (in)} \nabla^{- 2} \bar \Delta_{\rm com}^{\rm (in)} $. Using these relations and Eqs.~(\ref{eq:master})-(\ref{dbPhi}), we get immediately $\db u_{\rm M}^{\rm (in)} $, $\db \zeta_{\rm M}^{\rm (in)} $ and $\db \R^{\rm (in)} $.

\section{Conclusions}
We have explicitly computed, for the first time, at the perturbation level, the effect of pressure from baryons, radiation, neutrinos and any general equation of state with sound speed $c_s$ different from zero on the perturbations that arise from  single-field inflation which include (small) non-linearities in the metric. We have shown that $c_s, \Pi, \Gamma$ appear in the resulting perturbations thus, on scales/epochs when they are non zero these terms  induce breaking the single clock symmetry from single field inflation. 

The claims of Ref.~\cite{Tanaka_2011,dePutter:2015vga,Pajer:2013ana,Cabass:2016cgp} translated in asserting  that one could synchronize (observers) clocks to the inflaton one in patches of arbitrary size in the Universe. This in itself is very counter intuitive as General Relativity forbids this in the presence of curvature (gravity) as explicitely shown in Ref.~\cite{Rollo}. What we have shown here is that, in addition,  the presence of pressure also breaks the synchronization of clocks in patches smaller than the Jeans length. These patches can be very large  depending on the time chosen to set the initial conditions. As a general rule of thumb, patches must be larger than those where baryonic acoustic oscillations are measurable, which translates on about $150$ comoving Mpc for initial conditions set before recombination. For initial conditions set after recombination, the patches must be much larger than tens of Mpcs  and larger than neutrinos free streaming length in the presence of massive neutrinos.  This impossibility to synchronize clocks across separate patches opens up the possibility to measure the primordial non-Gaussianity from single-field inflation of the local shape which include a term $\sim$ $n_s-1$ and the General Relativity  $\sim -5/3$ term arising from modifications of the Poisson  equation at the horizon scale. Current and ongoing surveys like DESI, Spherex and LSST are forecasted to have enough statistical power to detect a non-gaussian contribution of the local type of an amplitude $f_{\rm NL}^{\rm loc}\sim 1$ ~\cite{Carbone,Karagiannis} making them well suited to access the inflationary signature. \\

\begin{acknowledgments}
DB and SM acknowledge partial financial support by ASI Grant No. 2016-24-H.0. Funding for this work was partially provided by  project PGC2018-098866- B-I00 MCIN/AEI/10.13039/501100011033 y FEDER ``Una manera de hacer Europa'', and the  ``Center of Excellence Maria de Maeztu 2020-2023'' award to the ICCUB (CEX2019- 000918-M funded by MCIN/AEI/10.13039/501100011033).  LV acknowledges support from the  European Union Horizon 2020 research and innovation program ERC (BePreSySe, grant agreement 725327). 
\end{acknowledgments}

\end{document}